\documentclass{article}

\usepackage{arxiv}

\usepackage[utf8]{inputenc} 
\usepackage[T1]{fontenc}    
\usepackage{hyperref}       
\usepackage{url}            
\usepackage{booktabs}       
\usepackage{amsfonts}       
\usepackage{nicefrac}       
\usepackage{microtype}      
\usepackage{lipsum}
\usepackage{graphicx}
\usepackage{threeparttable}
\usepackage{makecell}
\usepackage{diagbox}
\usepackage{xurl}

\title{COMPARATIVE ANALYSIS OF TECHNICAL AND LEGAL FRAMEWORKS OF VARIOUS NATIONAL DIGITAL IDENTITY SOLUTIONS}

\author{
Montassar Naghmouchi \\
  Samovar, Télécom SudParis \\
  Institut Polytechnique de Paris \\
  France, Palaiseau 91120 \\
  \texttt{montassar-bellah\_naghmouchi@telecom-sudparis.eu} \\
   \And
 Maryline Laurent \\
  Samovar, Télécom SudParis \\
  Institut Polytechnique de Paris \\
  France, Palaiseau 91120 \\
  \texttt{maryline.laurent@telecom-sudparis.eu} \\
  \And
Claire Levallois-Barth \\
Télécom Paris \\
Institut Mines-Télécom \\
France, Palaiseau 91120\\
  \texttt{claire.levallois@imt.fr} \\
  \And
  Nesrine Kaaniche \\
    Samovar, Télécom SudParis \\
  Institut Polytechnique de Paris \\
  France, Palaiseau 91120 \\
  \texttt{kaaniche.nesrine@telecom-sudparis.eu} \\}
  
\begin{document}
\maketitle
\begin{abstract}
 National digital identity systems have become a key requirement for easy access to online public services, specially during Covid-19. While many countries have adopted a national digital identity system, many are still in the process of establishing one.  
Through a comparative analysis of the technological and legal dimensions of a few selected national digital identity solutions currently being used in different countries, we highlight the diversity of technologies and architectures and the key role of the legal framework of a given digital identity solution. We also present several key issues related to the implementation of these solutions, how to ensure the State sovereignty over them, and how to strike the right balance between private sector and public sector needs.
This position paper aims to help policy makers, software developers and concerned users understand the challenges of designing, implementing and using a national digital identity management system and establishing a legal framework for digital identity management, including personal data protection measures.
The authors of this paper have a favorable position\footnote{The positions in this paper are those of the authors and are not these of any of the organizations that the authors are affiliated to.} \ for self-sovereign identity management systems that are based on Blockchain technology, and we believe they are the most suitable for national digital identity systems.
\end{abstract}


\section{Introduction}
\label{sec:1}
National digital identity systems are solutions endorsed by public authorities to provide digital identity attributes and/or digital identity documents, in order to identify and authenticate residents and citizens before accessing online public services. Some of the national digital identity solutions are also accepted by private sector to identify and authenticate their clients online.

A national digital identity solution is either created by the State, or created by private partners and backed and recognized by the State.

The emerging national digital identity solutions have opened up many discussions about the assurance level of online identity transactions because no in-person verification can be made between the service provider and the requester. Moreover, electronic trust services related to digital identity, such as electronic signatures and seals, are also made possible in most national identity solutions, opening the door for more online business transactions and bigger need for verifiability and accountability.

However, the legal recognition of these transactions must be addressed. Electronic trust services and online identity-related transactions are only as valuable and genuine as the legal framework that provides them with relevant purpose and validity. These legal frameworks also subject their associated digital identity solutions to data protection legislation, because of the sensitive nature of identity management data.
This means that studying a national identity solution requires studying the associated legal framework and  personal data protection legislation.
There are multiple research questions to cover within the theme of national digital identity systems. These are solutions that require a high level of availability, resilience and security. They are destined to be used by public authorities, citizens and residents, and allow access for sensitive and critical services. Liability (legal responsibility) of all involved entities requires legal recognition of the solution and of the transactions it allows for. This makes the legal frameworks associated to these solutions a crucial element.
Via comparison and study of different models and deployed solutions, we try to answer the following questions:\\
\textbf{(Q1) What are the architectural, technological, functional and governance choices that need to be made when designing and deploying a national digital identity solution?\\
(Q2) What are the key elements that need to be addressed in the legal framework of a given deployed technological solution? And how do legal framework and the technical solution affect each other? And what are the data protection measures these legal frameworks establish?\\
(Q3) What is the best identity model and infrastructure for a national digital identity system that ensures State sovereignty and empowers its users? How do we imagine this solution in terms of governance and ecosystem? And how can it benefit users, the public and the private sector?}

To answer these questions, we organize this paper as follows. The first section is a Background section where we present key definitions and technical overview of identity management models. The second section about National Digital Identity Solutions is dedicated to studying some digital identity solutions provided by public authorities in different countries, grouped by their identity management model of choice (to answer \textbf{Q1}). In the section Legal Frameworks for National Digital Identity Solutions, we are interested in two main legislation types: legislation that makes a digital identity official, i.e. recognized as legal by public authorities, and personal data protection legislation. These legislations are made to protect the online privacy of users and cover national digital identity solutions due to the sensitive nature of identity management data. The legal frameworks studied are the ones associated to the technical solutions presented in the section that precedes this (to answer \textbf{Q2}).
By studying and analysing the solutions, along with their legal frameworks, their governance and the different configurations such as the ecosystem and stakeholders, we can highlight the problems and issues facing any solution. It also enables us to point out the conditions for a solution's success.
We set out these conclusions and our own position as researchers in the field of identity and access management, law and privacy in the Authors' Position section. Our recommendations for the design and implementation of a national digital identity solution, as well as other considerations for protecting user privacy, ensuring state sovereignty over this digital governance instrument, and striking the right balance between public and private sector needs are also detailed in this section (in response to \textbf{Q3}).
Finally, we conclude our paper.

\section{Background}
\label{sec:2}

Identity management data and attributes have evolved over the years from simply coupling identifiers like usernames and emails with to-know authentication factors like passwords, to whole profiles characterized by an online presence, activities, content and other forms that help express and define “self-hood” \cite{who-am-i-online2}. This eventually meant the collection and processing of more personal data. Technologies and legislation regarding how these data are used, processed, and stored are also evolving.

However, the Internet was created without any standards for identifying users. As a matter of fact, anonymity is an important feature of the Internet, and it is what made online communities succeed as an alternative space capable of providing a multitude of identities, even fictive ones. It is important to preserve this kind of freedom over the internet.
In recent years, Internet freedom has become a major concern for users. A wide range of legislation worldwide reflect these concerns by providing a legal framework and directives for the processing, sharing, and storage of personal data by service providers.

Legal frameworks for online identity should preserve privacy, while allowing a form of accountability to prevent cyber-crime (e.g. fraud, hate speech, cyber-terrorism). Technological choices for building online identity solutions must align with legal frameworks. The same logic applies to national digital identity solutions, where users are provisioned with trusted digital identity attributes that can be used to perform the needed online transactions in a secure and private way, allowing user control, pseudonymity, and even anonymity in certain use-cases such as in health care.

\subsection{Definitions}
\subsubsection{Identity Management Data}
Identity management data, for a natural person (also referred to as a physical person or a data subject), are the multitude of data that are declared, computed, collected or issued across different online platforms that are directly or indirectly linked to a natural person, and used to manage their identity (identify, authenticate and authorize).
Identity management data range from identifiers, attributes and credentials that are self-certified or issued by a trusted party (public authority, university ...), to social media and online community profiles data, biometric data, identifiers created by or for the user and data resulted from internet activities. Identity management data also include authentication factors used by a user to prove that they are indeed the user described by these data, such as to-know factors (e.g., passwords and patterns), to-own factors (e.g., private keys, tokens, and physical objects), and to-be factors (e.g.,  fingerprints, facial recognition, and voice).

\subsubsection{Personal Data}
Personal data is defined as any information related to an identified or identifiable person\footnote{Article 4-1, GDPR.}.
This includes direct identifiers like the name, identification numbers and online identifiers, biometric data, social identity etc. It also includes any data that can be used to identify the person behind even indirectly.
In other legal frameworks, it is referred to as Personal Information or Personally Identifiable Information (PII)\footnote{OMB M-10-23 (Guidance for Agency Use of Third-Party Website and Applications), Appendix}.

eIDAS regulation defines Person Identification Data (PID) as a set of data enabling the identification of a natural or legal person\footnote{ Article 3-3, eIDAS regulation.}.
PID and any other data that can be used to identify a person directly or indirectly, mainly identity management data defined above, are considered personal data.

\subsubsection{Electronic Trust Services}
    Electronic trust services are electronic services related to the verification, creation and preservation of electronic signatures, seals, website certifications, electronic timestamps, and other similar services. The term was coined by the European parliament and the European Council first in the Electronic Signatures Directive 1999/93/EC\footnote{The term was initially Certification Service Provider} and later in the eIDAS regulation\footnote{Article 3-16, eIDAS regulation.}.
    The eIDAS regulation calls the provider of these electronic trust services a Trust Service Provider (TSP).
    If these electronic trust services meet additional eIDAS regulation requirements\footnote{Annexes I, II and III, eIDAS regulation.}, they are known as Qualified Trust Services\footnote{Article 3-17, eIDAS regulation.}, and the provider is knows as a qualified trust service provider (QTSP).
\subsubsection{Level of Assurance}
	The Level of Assurance (LoA) is the degree of confidence in a claimed digital identity or the certainty with which a claim to a particular identity during authentication can be trusted.
    eIDAS regulation defines 3 levels of assurance for electronic identification: low, substantial and high\footnote{Chapter 2, Article 8-2, eIDAS regulation.}.
    LoA is also known as IAL for Identity Assurance Level in other frameworks of reference.
    The US National Institute of Standards and Technologies (NIST), for example, defines different IAL for identity proofing, authentication, and federation \cite{NIST3}.
    Different LoA levels are required according to the variety of use cases.
\subsection{Identity Management Models}
\label{sec:1:2}
An identity-management model is a set of architectural and functional choices for managing identities in a given identity-management system. Traditionally, service providers used a \textbf{siloed identity management model} to identify and authenticate their users. Users have to sign up and sign in for each service separately and are obliged to keep their authentication information. 
From their end, service providers have to retain the identity management data and credentials of users. This means a more fragmented digital identity for users and an overhead for service providers tasked with the storage and security of their users’ data. Businesses have no way to obtain or share user credentials between them.

The \textbf{federated identity management model} solves this problem by centralizing most identity management data and credentials to be stored and managed by an identity provider (IdP) that connects the user directly with any service provider. Federated identity management is convenient for service providers, as it offloads the identity management tasks and costs, and it is also convenient for users who do not have to deal with scattered identity credentials and authentication material. However, the model presents a strong centralization and dependence on identity providers.

The \textbf{user-centric model} changes the game by placing the user at the center of the interaction between an identity provider and service providers. In \cite{idblog4}, Kim Cameron et al. define the user-centric approach as \textit{“structured so as to allow users to conceptualize, enumerate and control their relationships with other parties, including the flow of information.”}

Users are responsible for storing their identity management data and have the freedom to share them when needed with the service providers. 

The user-centric approach raises concerns regarding possible correlations and tracing of a user’s activities, as the IdP and service providers can collude to correlate users’ activities.
Although it enables the user to give consent and control who should share their identity management data, the user-centric model still relies on an identity provider to enroll the user and provide them with identifiers. The level of user autonomy is still weak.

The \textbf{self-sovereign identity model} (SSI) solves the autonomy and centralization problems of the typical user-centric model. Instead of relying on a centralized source of trust to authenticate users and on an identity provider to enroll and register them, the SSI model enables users to register themselves directly as many times as they wish on a decentralized platform like Blockchain, and enables them to contact and demand different issuers for different credentials and attestations/attributes to be linked to whatever identifier they want. Moreover, the SSI model enables users to create these credentials themselves, where they can self-assert claims and relationships about themselves and other users, such as club and group membership, personal capabilities, and other relevant claims. In general, SSI systems rely on open standards developed by different communities and work groups, like the Decentralized IDentifier (DID) and Verifiable Credentials (VC) standards and DID-Auth for authentication. Users interact with the SSI issuers and verifiers through an application called a wallet. Wallets allow users to request, obtain, store credentials they get from issuers and present them to verifiers. Wallets can also integrate other services like digital signatures, key and identifier generation  and even the capacity to create credentials that are either self-issued or issued by the users to other users, and the capacity to verify presented credentials. Wallet applications are not a compulsory component for an SSI system, since we can still have verifiable documents over a decentralized ledger similar to the Public Key Infrastructure (PKI) certificates model. In a matter of fact, issuers can still issue credentials that are rooted in Blockchain networks, and that are verifiable, and can simply transfer them to users by different means without relying on wallets. Users from their end can choose how to store and present their credentials. However, the process seems more complicated without a wallet application that provides such functionalities in a more secure way.

Table \ref{tab:1} compares the above four models on the basis of different features and security and privacy aspects.

\begin{table*}[t]
\centering
\setlength{\tabcolsep}{5pt}
    \caption{Comparison of identity management models on the basis of different performance and security and privacy factors.}
    \label{tab:1}
\begin{threeparttable}
\begin{tabular}{p{100pt}p{90pt}p{90pt}p{90pt}p{90pt}}
\hline
\backslashbox[39mm]{Features}{Identity model}
 &  Siloed & Federated & User-centric & SSI \\ 
\hline \\
Interoperability\tnote{1} & Low to None & Medium\tnote{2} &  High & High \\
\\
Control over identity management data and credentials & Service providers & Identity providers & Users & Users \\ 
\\
Registration Authorities & Applications & Identity providers & Identity providers & Autonomous registration \\ 

Privacy\tnote{3}
& High
(identity management data specific to each silo which mitigates correlations) & Low
(high risk of correlation across Service Providers) & Medium
(correlation still possible when IdPs and SPs collude) & High
(multiple identifiers, identity management data created, stored and controlled by users)
\\
\\
Decentralization & No & Yes but there is still a degree of centralization (limited number of IdPs per federation) & Yes & Yes, strong decentralization \\ \\
Service availability\tnote{4} & Medium 
(relies on the availability of the silo) & Medium
(relies on the availability of the IdP) & High & High (relies on the availability of the distributed ledger) \\ \\
Common standards and protocols & TLS/SSL, HTTPS & SAML, OAuth, OpenID Connect & OpenID, SpaceCard & 
DID, DID-Auth, VC, DID-Communication, Blockchain\\ \\

User Autonomy &
Low 
(identity management data provided and managed by the silo) &
Low
(identity management data provided and managed by IdPs) &
Medium
(identity management data provided by IdPs but controlled by users) &
High
(the users generate, publish and manage their own identity management data) \\ \\

Architecture
& 
    \includegraphics[width=30mm]{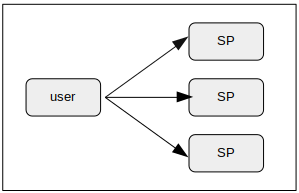}

& 

    \includegraphics[width=30mm]{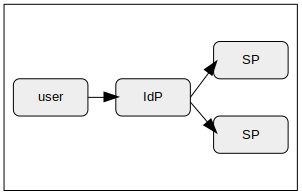}
&

    \includegraphics[width=30mm]{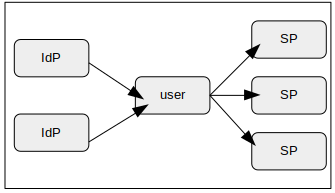}
&

    \includegraphics[width=30mm]{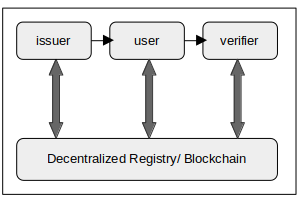}
\\
\hline
    \end{tabular}
\begin{tablenotes}
\item[1] Use of identity attributes across different systems
\item[2] Only between federation members
\item[3] Measured according to the level of possible correlations among entities to link identity management data, and the number of entities able to link them
\item[4] Availability of the identity and access management service

\end{tablenotes}
\end{threeparttable}
\end{table*}

\section{National Digital Identity Solutions}
\label{sec:3}

Digital identity systems vary from one State to another, from Single Sign-On (SSO) authentication systems that enable citizens to identify and authenticate themselves with different online public services, to fully mobile wallets containing digital versions of official credentials - issued by public authorities or on behalf of public authorities.

The different design choices on one hand, coupled with governance choices, are what makes national digital identity solutions very different in terms of capabilities, levels of assurance in the user's identity, what services they allow their users to access and how much do they appeal to public use or adoption by the private sector. We proceed by studying a few examples of these solutions grouped by their identity management model that we presented in the previous section. 

\subsection{Federated Identity Management Solutions}
\label{sec:3:1}
Multiple national digital identity solutions around the world follow a federated identity model. Notably, two solutions that are notified within the "electronic Identification, Authentication, and trust services" (eIDAS regulation) regulation and are the French FranceConnect/FranceConnect+\footnote{To the moment of writing this article, only FranceConnect is rolled out and currently being used. Thus we focus on FranceConnect.} and the British UK Verify - at least before Brexit. Both solutions follow a federated identity management model and are used by users to access online public services. However, while FranceConnect has over 30M users by October 2021 \cite{Franceconnect-users5} and is still operational, UK Verify failed to attract British users, gaining only about 3.6M users at its peak \cite{Verify-Audit6} and was eventually decommissioned by 2020, only to be extended during Covid-19. Differences in business logic, technologies, and legal frameworks between the two States may explain these different outcomes.

For instance, the manner in which identity providers (IdPs) are designated and chosen for each solution is a major difference. France relied on its own public authorities to work as IdPs for FranceConnect since most of them already have an identity solution - e.g. the ministry of finance (impots.gouv), the French Social security (Ameli), the French post office (Identité Numérique La Poste) and "France Identité" as an IdP provided by the ministry of interior, ministry of justice and other public authorities.

On the other hand, the UK has relied exclusively on private sector companies as identity providers, mainly Experian, Verizon, Barclays, GB Group Digidentity and Post Office \cite{UKV-IDP7}. Currently, all have withdrawn, except for the Post Office and Digidentity. This had major drawbacks on the solution. For example, Experian alone had around 2M registered users, it was the biggest IdP for Verify, and when Experian opted out from their role in the UK Verify, they simply asked their users to re-register with another IdP, with a total lack of interoperability and availability.

While FranceConnect only offers access to online public services, UK Verify enables using services from private sector entities using Open Identity Exchange (OIX)\footnote{ Most IdPs for UK Verify are members of the OIX, mainly Barclays, Experian, GBG and Digidentity \cite{OIX-members8}}. However, it is planned that FranceConnect+ will enable such services.

Another major difference between FranceConnect and UK Verify is that FranceConnect had no competitors supported by the public administrations. Although public administrations had their own digital identity solutions and identity management data, they came together within FranceConnect as identity providers for the federated solution. That was not the case for UK Verify, which competitor - “Gateway Connect” developed by HM Revenue and Customs (HMRC) - gained over 16M users\footnote{Gateway Connect is based on the Government Connect solution that was supposed to be replaced with UK Verify. Instead, HMRC developed its own solution that uses the existing Government Gateway identifiers.}, and already enabled users having an account to authenticate themselves and access around 123 online public services, a number far greater than the public services that UK Verify enables users to access.

Moreover, HMRC refused to integrate UK Verify as an authentication method for their services \cite{HMRC9}. 
This was also the case in the National Health Service (NHS) of England, who found that UK Verify was not suitable from a security and privacy perspective  for authenticating patients and allowing them to access health data. NHS patients who have tested the solution were highly skeptical of banks and private entities providing their identities and intervening in their health data \cite{NHS10}.

\begin{figure}[h]
    \centering
    \includegraphics[width=80mm]{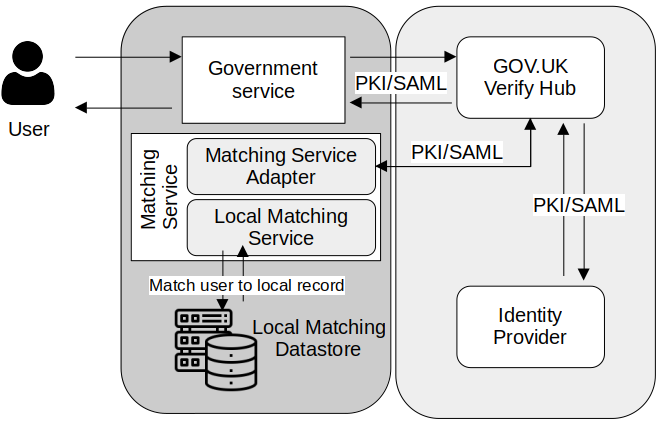}
    \caption{UK Verify general architecture and matching flow \cite{UKV11}}
    \label{fig:uk-verify}
\end{figure}

Even from legal and social points of view, the grounds for a digital identity system differ between the two countries. France still has national identity cards and centralized databases containing the identity management data of natural persons residing in France (RNIPP), which facilitate the identification and authentication of individuals. For example, the RNIPP database was consulted to  authenticate the presented identity management data and to remove redundant identity profiles from FranceConnect. On the other hand, the UK does not have national identity cards or centralized identity databases, as data are fragmented across different agencies and authority branches, making it harder to authenticate and de-duplicate identity management data at the national level. That is why, as depicted in Figure \ref{fig:uk-verify}, the UK Verify solution uses a Local Matching Service that is hosted on the organization’s infrastructure (for example, each public authority should have its own identity database) and it only provides a Matching Service Adapter to help match identity management data provided by an IdP to locally existing data records. The matching policy is also left to the discretion of organizations which can deal with single, multiple or no matches. 

On a technical level, the two solutions integrate different identity federation protocols even though they follow the same identity model. FranceConnect uses the Open ID Connect (OIDC) protocol, which is more recent than the SAML protocol used by UK Verify. In matter of fact, OIDC is the third generation of the Open ID protocol, and was published in 2014, while the latest version of SAML, SAML 2.0, dates back to 2005.

OIDC uses JSON over the RESTful API to transfer identity management data between IdPs and service providers, whereas SAML uses XML over HTTP and SOAP to transfer identity management data. Communication between IdPs and service providers in SAML is secured using encrypted, digitally signed certificates, whereas in OIDC, encrypted, digitally signed tokens are used.

OIDC is easier to consume and lighter to use with APIs, whereas SAML does not work well with APIs and is too heavy. Moreover, OIDC can also be used for mobile applications, unlike SAML, because its reliance on XML is difficult to integrate into the mobile world. These fundamental differences between the chosen protocols were also inherited by the  two solutions, and may explain, among other factors mentioned above, why FranceConnect was more successful than UK Verify and why it is more appealing to users.

\subsection{User-Centric Solutions}

User-centric national identity solutions usually revolve around an official digital identity document or identifiers held on a mobile application or a physical card equipped with a smart chip used via a card-reader or a compatible equipment, which the user employs with various service providers both online and offline public services, and other private sector services that accept a State-authenticated digital identity.

India’s Aadhaar, Estonia’s e-ID and Singapore’s Singpass are examples of user-centric national identity solutions. Although all three solutions follow the same identity model at an architectural level, there are significant differences.
For example, Aadhaar and Singpass are highly centralized in terms of registration. The only identity provider is the State, and registrars are either public authorities or third parties working on their behalf. The acquired identity management data are held by the public authorities in a centralized way. On the other hand, Estonia’s e-ID  handles identity management data in a decentralized way using the X-Road platform and a Blockchain platform.
\subsubsection{India’s Aadhaar}
For Aadhaar, India has enrollment centers responsible for obtaining the biometric and demographic data of an individual and enrolling it in the system by providing a unique permanent 12-digit identifier. The system has already issued 1.3 billion identifiers to users\footnote{As of March 2021 according to Unique Identification Authority of India (UIDAI) Annual Report 2020-21} where each user has one unique identifier, and the deduplication process - which is the removal of redundant identifiers or the detection of subjects having more than one identifier - is done by consulting a biometric database called the Central Identities Data Repository (CIDR) owned and centralized by the public authorities. Three Automatic Biometric Identification System (ABIS) from three different vendors are used to detect redundant biometrics by comparing incoming CIDR reference and anonymized biometrics to their own \cite{ABIS12}. Aadhaar identifiers and cards allow holders to identify and authenticate themselves to access online and offline services such as banking, sim cards, social pensions and rations, and other online public services. Aadhaar also enables digital signatures with the same legal value as handwritten signatures.

However, the world’s largest digital identity solution has significant privacy and accuracy concerns. For example, relying on a biometric database to deduplicate identifiers is problematic and cannot guarantee the correct deduplication. Between 2010 and 2019, 475,000 identifiers were canceled, with an average of approximately 145 duplicate identifiers per day \cite{aadhaar issues13}.
In addition to the issues of proper operation, Aadhaar has several privacy and confidentiality issues. For instance, having a unique global identifier using the same identifier for all online transactions means that the CIDR can correlate the user's activity and has a complete history of who accessed what.

Privacy International published a very critical report of Aadhaar \cite{PIAadhaar14}, and cited the impossibility of making efficient comparisons to deduplicate identifiers from a biometric database of the size of the Indian population (approximately 1.4 billion).
The report also notes data breaches and leaks in the system. There are also important violations of user consent, where the user is obliged - in practice - to enroll and have an identifier even if it is not legally mandatory. Aadhaar identifiers are also linked to bank accounts and health records, further compounding the problems and compromising solutions.

\begin{figure*}[h]
    \centering
    \includegraphics[width=120mm]{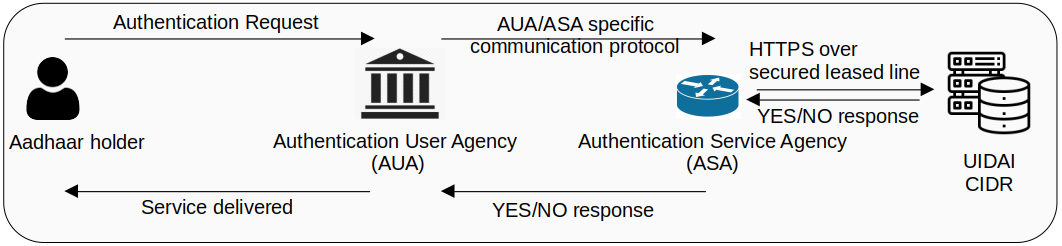}
    \caption{Aadhaar - user enrollment and authentication process ; The CIDR is a crucial component \cite{PIAadhaar14}}
    \label{fig:aadhaar}
\end{figure*}
As Figure \ref{fig:aadhaar} shows, the Aadhaar system relies heavily on the CIDR biometric database, which calls into question the performance and accuracy of the system as a whole because the verification method, that is, comparing the submitted data with the CIDR data, is not reliable in itself. The heavy reliance on the CIDR database and biometric and demographic data for deduplication makes it mandatory for users to give up their biometric data during enrollment to be issued an Aadhaar identifier. This technical obligation to collect biometrics for correct functioning exerts a heavy toll on the Aadhaar solution because of the sensitivity of the data, extra security measures, and overhead needed to store and transport them. 

Aadhaar also raised transparency issues as no public consultations took place during the design of the solution. Moreover, unlike all the solutions analyzed in this section, Aadhaar is a proprietary solution that is not open-source. 

\subsubsection{Singapore’s Singpass}
Singapore’s Singpass is also based on a authority-owned biometric database. 
It uses an API called "Identiface" to authenticate users through online facial recognition and allows them to access online public and private sector services, as well as electronic signatures. Singpass comes with a mobile app in the form of a wallet storing digital identity documents, such as drivers’ licenses, national ID cards, diplomas, etc. These digital documents are in the form of digital certificates issued by the National Certification Authority of Singapore. The solution relies on a PKI for digital signatures and digital certificates issued to users mainly by the centralized national registration identity card (NRIC) of Singapore. 

Singpass provides two main services for citizens and businesses: MyInfo which uses the latest data on individuals and businesses from various public authorities to automatically fill out forms and exchange these data between the involved parties, and Verify which is used to verify the data without any physical documents. 4.5 million persons use the Singpass system (around 82\% of the Singaporean population and 97\% of citizens and residents above 15 years old) \cite{singpassUserStat15}, and about 78\% use the Singpass mobile application.

Like Aadhaar, Singpass is government-owned, managed, and maintained by the Government Technology Agency GovTech, but it is an open source solution \cite{singpassGithub16} and has received less criticism about transparency and openness.

The use of facial verification API in a national identity scheme is the first, which comes with privacy concerns, fear of mass surveillance, and intrusiveness. Users consent to the use of their biometric data for authentication. The use of mobile applications and mobile identity documents is optional because users can use Singpass as a website to access services using their physical identity documents and the identifiers present on them. Providing many alternatives to access online services is always crucial for digital identity schemes because if a national identity scheme is left to be the only way a user can access services, it is no longer an optional solution, even if it is stated to be one, which is the case with Aadhaar.

\subsubsection{Estonia's e-ID}
In terms of digital identity infrastructure, Estonia is arguably the most developed national digital identity system. Estonia launched e-governance and e-banking services in 1996, an e-tax for online tax payments in 2000, and the national digital identity program e-ID in 2002. The national identity card is mandatory in Estonia from the age of 15; 99\% of Estonians have national ID cards, that are equipped with an electronic chip.

Estonia uses a PKI system to power the e-ID infrastructure: a private key is generated and stored on the card’s electronic chip, the identifier on the card is used as the public key, and the public key can be retrieved using LDAP from the public directories.
The card and the associated software called “DigiDoc” can be used to authenticate the cardholder online using PIN1, to electronically sign documents using PIN2, to encrypt data and exchange them using the keys associated with the card, and to access public and private online services.

The Police and Border Guard Board (PBGB) is the authority responsible for the creation and supervision of both physical and digital identity documents to Estonians.

The shortcomings of e-ID cards are that users need a card reader to use their identity cards with their computers. Estonia launched its Mobile ID program in 2007 with the country’s major mobile operators. Today, approximately 19\% of Estonians use Mobile ID, which is a special SIM card equipped with applications for digital signatures and authentication and private keys stored on the chips, enabling the same functionalities of an Estonian e-ID card without relying on a card reader.

To ensure the availability, confidentiality, and integrity of citizens’ identity management data, Estonia uses a Keyless Signature Infrastructure called the KSI Blockchain \cite{KSI17}. The KSI Blockchain is used as part of a data exchange layer to timestamp transactions via an X-road platform, which is used by both public and private entities in Estonia to exchange data. In addition, Estonia has “Data Embassies,” the first of which is in Luxembourg. These embassies are data centers, servers, and cloud instances capable of running all online services and providing backup data in any situation to ensure service continuity. This solution is crucial to ensure availability because Estonia has a paperless policy and heavily relies on a digital infrastructure that was attacked on several occasions, the most critical of which was in 2007 \cite{EstoniaAttack18}.

Despite being advanced and well established, the Estonian national digital identity scheme experienced several troubles. For example, in 2011, faulty cards were delivered to users, and it took the Estonian authorities over nine months to publicly and transparently address the incidents. A similar problem occurred in 2017, where the same private key was stored on multiple cards by a private contractor \cite{EstoniaInfineon19}. The latter incident was immediately communicated by the public authorities, but it took time to find and resolve the problems, which involved approximately 750,000 cards at the time.

Estonia contracts private companies to build various technological components of its digital identity infrastructure, which means an extra layer of abstraction between the public authorities and the digital identity services it provides. This extra layer can prevent public authorities from overseeing the services correctly and efficiently.

\begin{figure}[h]
    \centering
    \includegraphics[width=60mm]{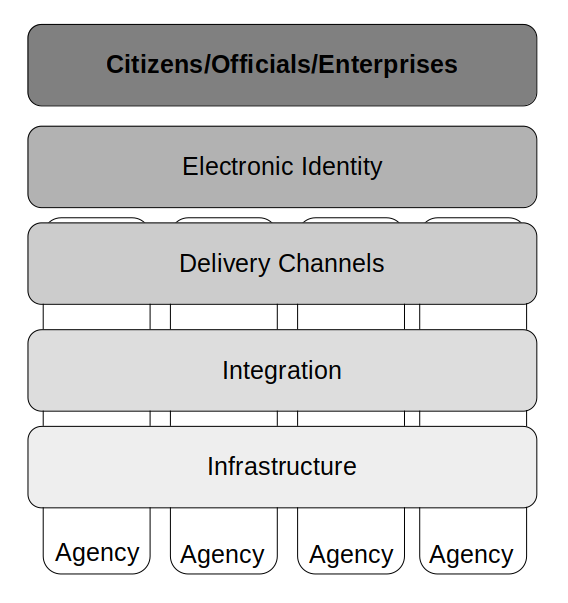}
    \caption{Layers enabling e-governance and the digital transformation in Estonia \cite{Estonia-eGov20} }
    \label{fig:estonia-layers}
\end{figure}

The digital identity layer is a pillar of e-governance. It enables all public and private services and allows secure data and document exchanges between all involved parties, as depicted in Figure \ref{fig:estonia-layers}.

\subsubsection{Summary of user-centric solutions}
User-centric identity solutions still present privacy issues, and even if users control their identity management data, identity and service providers can still correlate user's activities and relationships. The centralized sources of truth – for authentication and  deduplication – on which these solutions rely, such as centralized or decentralized databases held by the public authorities or identity providers, still introduce design flaws and opportunities for hackers to target these databases, thereby threatening data confidentiality and integrity.

In matter of fact, both user-centric and federated identities have conceptual drawbacks. Both models rely on identity providers, and whether the identity provider is from the private or public sector, the immense levels of power that the provider has over the user’s identity make the relationship completely asymmetric. This is inappropriate for the users, since IdPs can revoke identities, delete identity management data, or simply stop maintaining the systems, as was the case with UK Verify.

Rethinking how digital identity should be designed has become crucial, as it is a way to empower individuals instead of aggravating mass-surveillance and allowing private organizations to merchandise personal data.

\subsection{Self-Sovereign Identity Solutions}
Powered by a distributed ledger, generally a Blockchain, Self-Sovereign Identity (SSI) systems are increasingly being adopted in identity systems that require a high level of decentralization, interoperability, and shared governance. Companies such as Evernym and organizations such as Sovrin are already providing Blockchain-based digital identity infrastructures and services for various use cases \cite{SSI-Evernym-usecases21} centered on verifiable credentials and their exchange.

The European Self-Sovereign Identity Framework (ESSIF)\footnote{Note: while ESSIF itself is a framework and not a single monolithic solution, for simplicity we refer to it as one.} is perhaps one of the most interesting, promising and challenging implementations of the SSI model on a big scale.

The ESSIF aspires to provide a technical framework in which different actors work together to solve one of Europe’s digital challenges: a digital identity solution that is in accordance with the legal framework defined by eIDAS regulation - which is currently being discussed and evolving to eIDAS 2.0, refer to \textbf{eIDAS regulation (EU)} in the next section about legal frameworks - and trusted in the European Union. 

\begin{figure*}[h]
    \centering
    \includegraphics[width=120mm]{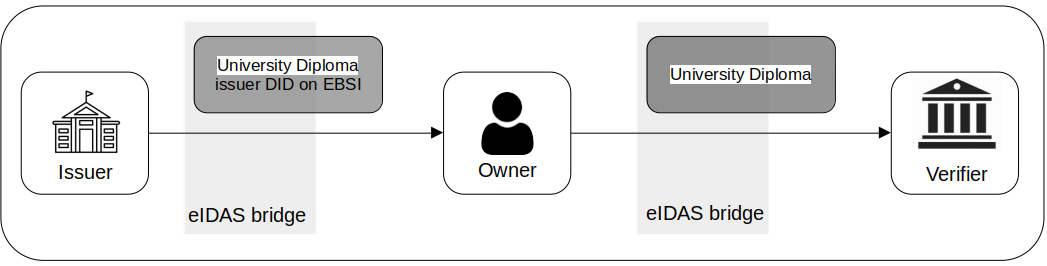}
    \caption{SSI over eIDAS bridge to have recognizable VC across the EU ; VCs are linked to DIDs stored on the EBSI Blockchain \cite{EC-eIDAS22}}
    \label{fig:eidas}
\end{figure*}

The EU identity program consists of three major building blocks: the European Blockchain Services Infrastructure (EBSI) providing a consortium Blockchain between EU members; eIDAS nodes for interoperability between national infrastructures and administrations, including national identity infrastructures; and the ESSIF framework that provides users with an SSI application used for digital identity services. Figure \ref{fig:eidas} illustrates this.

\subsubsection{ESSIF context and Scope}
\begin{figure}
    \centering
    \includegraphics[width=75mm]{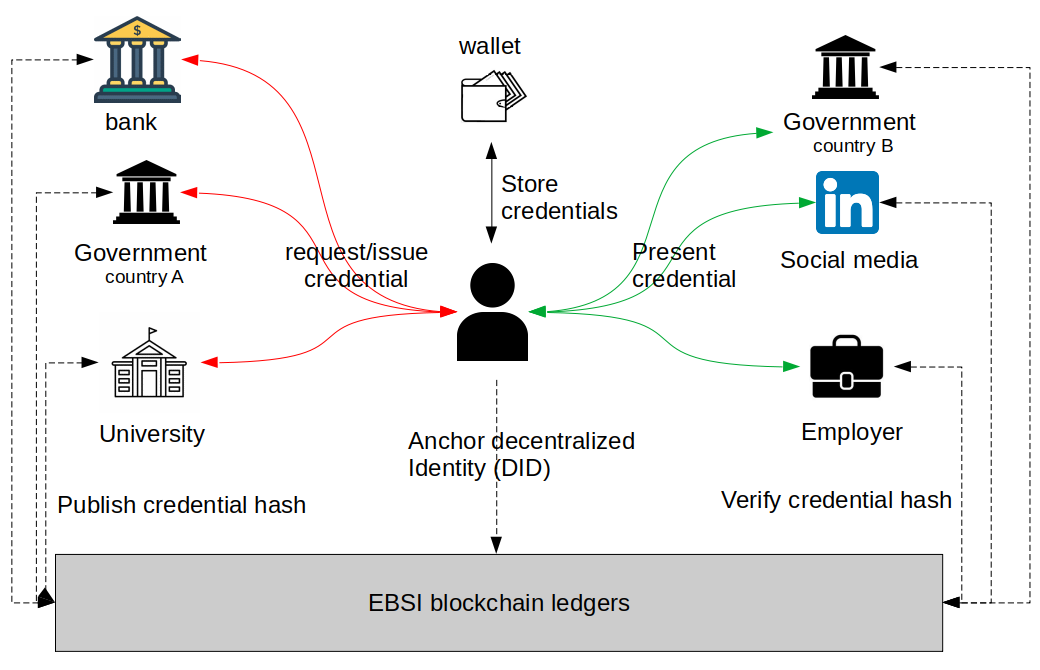}
    \caption{High level architecture of ESSIF}
    \label{fig:essif}
\end{figure}

The ESSIF defines an ecosystem for the European SSI and a trust framework that allows natural persons to be identified, authenticated, and exchange verifiable credentials, which are digitally signed attestations of attributes about the subject issued by a trusted entity, such as a university or a public authority. These credentials are designed to be cryptographically verifiable in terms of ownership and authorship, and support different privacy enhancing techniques, such as zero-knowledge proofs and selective disclosure, to guarantee a high level of privacy.
The first use case of ESSIF is to create digital diplomas as verifiable credentials issued by universities to their alumni. These credentials are to be verified by potential employers and stakeholders across the EU. Figure \ref{fig:essif} is an overview of ESSIF that shows potential stakeholders and transactions between them.

\subsubsection{ESSIF Governance and Ecosystem}
For a project on this size to work, the ecosystem must be carefully defined. The ESSIF is designed as a large framework, where multiple SSI solutions and services are recognized cross-border between EU Member States. ESSIF is one of the first use cases of the EBSI Blockchain.

EBSI blockchain is specifically used by ESSIF for publishing decentralized identifiers, public keys and credential related metadata such as schemas and revocation status. In addition, the ESSIF maintains public lists for trusted issuers and trusted service providers to ensure greater trust in the framework. ESSIF does not aim to build or maintain a wallet for EU citizens but rather to define a solid framework for different wallet providers and electronic trust services to work within. 

\subsubsection{ESSIF/EU ID Wallet}
The EBSI project includes a wallet, but only for testing purposes, and is not  or intended to be publicly used. ESSIF, as a framework, as well as the work of the eIDAS Expert Group on EUDI wallet specifications \cite{ARF-eIDAS23}, propose architectures, standards, and requirements for wallet applications, rather than providing one, because other actors, from both public and private, are supposed to provide their own wallet solutions to users.

These entities are called “wallet providers” and will be certified by a conformity assessment body and under the supervision of a supervisory body. Wallets can be both online and offline, with different requirements for each type, and must allow users to register themselves by publishing their DID on an EBSI ledger and to communicate with various issuers and trust service providers to obtain verifiable credentials and perform normal and qualified electronic signatures.

The capabilities and requirements of the wallets, as defined by the eIDAS Expert Group, go beyond the scope of ESSIF for some functionalities such as digital signatures and other electronic trust services, but in general align in terms of wallet identity requirements and standards, since ESSIF must align with and comply with the eIDAS regulation. Overall, although the EUDI wallet does not mention any standards to be used in terms of identifiers and credentials, the ESSIF does.

The ESSIF encourages the use of the DID standard as Blockchain identifiers and VC standard as credentials to be used by different interacting parties. EBSI even propose their own DID method called DID EBSI \cite{DID-EBSI24}. Users can register as many DIDs as they want and can use different DIDs to interact with different issuers that have different levels of assurance based on the type of VC they provide. Different VCs obtained from different issuers can be linked to different DIDs or the same DID depending on what the user wants and on the use-case.

After a DID is registered, the legal implication of using that identifier and the associated credentials are still in discussion since different EU Member States have different notions and laws regarding unique identifiers is and if a DID is considered a unique identifier in this context or not. 
However, a DID can be legally and officially linked to the identified subject when needed by having an authority to verify the legal identity of the subject and issue a credential linking that identity to the DID. Moreover, a VC may contain the digital signature or digital seal of a qualified trust service provider, which adds value and LoA to digital documents or claims issued in the form of a VC \cite{eIDAS-SSI25}.

\subsection{Section Summary}
Implementing a national digital identity system presents challenges from one State to another. Some States that already have uniform (centralized or decentralized) databases of citizens and resident identities, such as France (RNIPP, INSEE registry), Estonia, Singapore, and India, have been more successful in implementing digital identity systems with higher levels of trust in the authenticated identity. In contrast, countries with more fragmented digital identity databases and without national identity schemes such as the national identity card, e.g. the United States and the United Kingdom, have had more difficulty implementing such systems. National digital identity projects follow different identity models, depending on legal frameworks and the ability of different existing identity infrastructures to coexist and be used in newer digital solutions, as summarized in Table \ref{tab:2}.

On the other hand, we have SSI solutions emerging all over the world, such as in New South Wales (Australia) \cite{NSW26} and British Columbia (Canada) \cite{BCVC27} and ESSIF.

ESSIF starts with a well-defined use case: the cross-border recognition of educational credentials such as diplomas and student IDs. This first use case will be used to prove that a European SSI platform is an essential step towards a single European market. Providing full administrative interoperability via a cross-border digital identity will enable European citizens to access services in all EU Member States in the same way as they do in their own country.

\begin{table*}
\centering
\setlength{\tabcolsep}{7pt}
    \caption{Comparison of several national digital ID systems.}
    \label{tab:2}
\begin{threeparttable}
\begin{tabular}{p{75pt}p{50pt}p{50pt}p{50pt}p{50pt}p{50pt}p{50pt}}
\hline
\backslashbox[30mm]{Properties}{System} &  
FranceConnect&
UK Verify &
Aadhaar & 
eID Estonia &
Singpass &
ESSIF 
\\ 
\hline
\\
Decentralized identity database &
No &
Yes &
No &
Yes &
No &
Yes \\ \\
Identity model &
Federated &
Federated &
User-Centric &
User-Centric &
User-Centric &
SSI \\ \\
 IdP &
Public authorities,
Private companies &
Private companies &
Public authorities &
Public authorities &
Public authorities &
None \\ \\
Services from private sector &
No
&
Yes &
Yes &
Yes &
Yes &
Yes \\ \\
Electronic signatures &
No &
No &
Yes &
Yes &
Yes &
Yes \\ \\
Interoperability &
Yes\tnote{1}
 &
Yes
 &
No &
Yes
 &
No &
Yes \\ \\
Standards, protocols and technologies &
OpenID, OAuth-2 &
SAML, PKI &
ABIS &
KSI Blockchain, X-road, PKI, SIM cards, electronic chips &
Facial recognition &
DID, VC, Blockchain, DID Auth
\\ \\
Deduplication method &
Cross-checking the RNIPP database &
Local Matching &
Biometrics &
Not specified,
but supervised by the PBGB &
Biometrics &
No single identifer 
\\ \\
Legal framework &
eIDAS regulation, GDPR,  the decree of November’s 8th 2018 &
eIDAS regulation, GDPR &
Aadhaar Act 2016 &
eIDAS regulation, GDPR, the Identity Documents Act &
NDI\tnote{2}  initiative &
eIDAS regulation, GDPR
\\ \\
Openness &
Open source &
Open source &
Proprietary &
Open source &
Open source &
Open source
\\ \\
\hline
    \end{tabular}
\begin{tablenotes}
\item[1] Via eIDAS node, the same goes for UK Verify, Estonia's eID, ESSIF and any other European national identity scheme that is eIDAS notified.
\item[2] National Digital Identity
\end{tablenotes}
\end{threeparttable}
\end{table*}

From experience and by referring to different incidents, such as: (i) the scrapping of the ID Cards in the UK in 2011 (ii) the SAFARI affair in France in 1972-1974 where the public authorities attempted to interconnect identity files and data held in different French administrations and authorities using the INSEE number and (iii) the United States where President Bush’s attempt to create a national ID card failed for fear of a lack of transparency and abuse of the system in the absence of a clear legal framework, we can conclude that if a national digital identity system is to work, it must avoid repeating mistakes like lack of transparency and the absence of a legal framework that guarantees the proper use of personal data, privacy protection, and trust. In the next section, we conduct a comparative study of Legal frameworks related to the national digital identity solutions presented above.

\section{Legal Frameworks for National Digital Identity Solutions and Personal Data Protection Legislation}
\label{sec:4}
Legal frameworks that are established to regulate how digital identity solutions and electronic trust services are deployed and used. They add legal recognition to identity-related transactions like electronic signatures and seals. These frameworks also include personal data protection measures.

\subsection{Legal Frameworks For National Digital Identity Solutions}
\label{sec:4:1}

National identity solutions are often created or launched by legislation that makes them possible within a given legal framework. The legal framework defines governance for a solution by creating or designating a regulatory control body. The framework should also define how the data collected for these solutions are stored, processed, and used, especially personal data.
Generally speaking, depending on the State, the framework either makes the solution subject to existing data protection legislation or establishes specific rules for personal data protection.

\subsubsection{eIDAS regulation (EU)}
\label{sec:4:1:1}
The eIDAS regulation is a EU regulation adopted in 2014 and went into force in 2016. The regulation addresses the electronic identification cross-border between EU Member States, by virtue of mutual recognition of eIDAS notified digital identity schemes for high and substantial LoA. Private sector services can also make use of eIDAS notified schemes but they are not under any obligation.
Overall, the regulation aims to harmonize and leverage standards for better interoperability between public digital infrastructures and services in the EU.
The regulation currently has 5 electronic trust services in scope: electronic signatures, electronic seals, electronic time stamps, electronic registered delivery services and website authentication.
The regulation differentiates between qualified and non-qualified trust services and trust service providers. A qualified trust service provider providing qualified electronic signatures or qualified digital certificates, for example, must comply with additional requirements in eIDAS regulation\footnote{Article 24 and Annexes II,III,IV, eIDAS regulation.}.

There are several eIDAS-notified national digital identity schemes with different LoA, allowing for different use cases depending on the required LoA. While eIDAS regulation does not require States to have an eID solution or to notify one, it does require Member States to accept the notified eID schemes\footnote{Article 6, eIDAS regulation.} of other EU States for accessing public services if these schemes are of LoA substantial or high. 
This cross-border recognition of eID enables EU citizens and residents to access public and private services of other Member States. EU-wide recognition of eIDAS notified solutions started in 2018.\\
The eIDAS Regulation also addresses personal data processing and protection, stating that it must be carried out in accordance with Directive 95/46/EC, which is now effectively replaced by the GDPR. Supervisory bodies are appointed by Member States to oversee the implementation of the regulation and cooperate with data protection authorities (required by the GDPR).

eIDAS regulation aims to be technology-neutral, leaving Member States free to choose whatever technology they wish, as long as the requirements of the regulation are met.

eIDAS nodes, also known as eIDAS bridges \cite{eIDAS-node28}, is a technical solution which operate in Member States to ensure interoperability between national infrastructures and to ensure that authentication points in each Member State can recognize and verify presented electronic proofs  of identity or identity-related transactions. Projects such as the ESSIF and EBSI are also based on this regulation, which means that eIDAS regulation is a very important legal framework for the European digital market.

According to an evaluation study of the eIDAS regulation published by the European Commission in 2019 \cite{EC-study29}, only 14\% of key public-services are available cross-border via eIDAS notified solutions, while the majority of public service providers do not support cross-boarder authentication. This low performance compared to eIDAS aspirations have lead to the current eIDAS 2.0 discussions that aim to provide a EU digital identity wallet like the one discussed in ESSIF section previously. In matter of fact, eIDAS 2.0 have seen a paradigm shift compared to the previous version. In matter of fact, in 2021 the European Commission have proposed a framework for a European digital identity within eIDAS, based on an identity wallet \cite{EC-ID30}. This framework will amend eIDAS to eIDAS 2.0.
The eIDAS 2.0 will make credential issued from a public authority issuer to holder wallet to have the same legal effects of a qualified certificate of attributes, among others.

\subsubsection{The Decree of November’s 8th 2018 (France)}
\label{sec:4:1:2}
The French national digital identity scheme, FranceConnect, started in 2014 and was regulated in 2018 by the decree of November 8th. FranceConnect is the successor of “Idenum,” a project launched in 2009 with the goal of providing a digital identity infrastructure to facilitate access to online public services. Although Idenum’s initial work saw several organizations (SFR, La Poste ..) partnering, the stakeholders had their own digital identity projects in parallel. These competing solutions led to the failure of the Idenum initiative, especially because of the FranceConnect project that was launched in 2014 and promoted a single national information system to deliver public services based on August 2014 decree, No. 2014-879. 

In 2016, the solution went alive, and in November 2018, a decree was published to provide it with a legal background and placing it under the supervison of interministerial digital management DINUM (Direction Interministérielle du NUMérique). \\
In 2021, FranceConnect was eIDAS-notified as the national identity system of France. In matter of fact, it was notified under the name "FranceConnect+" with La Poste as an identity provider under DINUM to have the LoA substantial.

The decree of 8 November 2018\cite{AR8Nov} defines the purpose, context, and scope of FranceConnect. The solution is described as a federated identity solution that is used to facilitate access to public services and online services provided by private organizations and to secure data exchange between public administrations. Because it is eIDAS notified, FranceConnect+ also permits users to access digital services in other EU Member States.

The decree also defines the personal data collected and processed for the legitimate operations of FranceConnect. Mandatory data include the last name, first name, date of birth, place of birth, email address, and other identifiers generated by the system. A unique technical alias is obtained from the hash of the personal data and used to de-duplicate identities with reference to the RNIPP. Optional data include the username and cell phone number of the user. Other data are collected for traceability, for example, IP addresses and ports, connection timestamps, and web tokens issued by identity providers.

Furthermore, the categories of data processing, the data controllers in charge of the data processing and the manner in which the data are stored and protected are specified. This includes defining recipients of personal data and for how long personal data can be stored.
The text also addresses the right to access and the right to rectification and deletion. By the decree 2021-1538, users can consult and access personal data held by different public administrations. This can be done via a public service called MonFranceConnect (in Beta version so far).
FranceConnect must also comply with a broader regulatory context, including the French Act 78-17 supervised by the CNIL (Commission Nationale de l'Informatique et des Libertés - National Commission for Information and Liberties), the GDPR, and the eIDAS regulation.
Through the decree of November 8th, the legal aim and finality of FranceConnect, as well as the scope of use, are clearly defined and are intended to set limits of use to be respected by the public administrations. Another notable point is that the decree freezes the identity management model because it specifically references the federated identity solution.

\subsubsection{Identity Related Legislation in the UK}
\label{sec:4:1:3}
Identity related legislation in the UK is a bit complex. For starters, the existence of national identity cards has been a subject of conflict since they were introduced by the Identity Document Act 2006 along with the National Identity Register that contains information about identities. This registry is similar to the French RNIPP discussed above in the comparison between UK Verify and FranceConnect. By 2010, the Act of 2006 was repealed by the Identity Document Act of 2010 that cancels ID cards and forces the destruction of all information recorded in the National Identity Register. UK Verify, started by the Government Digital Services (GDS) as early as 2011, was officially launched in 2016 and eIDAS notified in 2018, in the context of the GDS tasks of provision of online public services. The solution was recognized on a European level, but on the national level, other public authorities had their own solutions going without any attempts to unify the solutions or integrate the UK Verify within them.

The after-phase of UK Verify on the other hand seems more organized and defined: the UK has established the Office for Digital Identities and Attributes (ODIA) in 2022 to supervise security and privacy for digital ID solutions. Moreover, the new Data Protection and Digital Information (No. 2) Bill draft 2023 establishes in its second part "Digital Verification Services" a trust framework\footnote{ Section 47, Data Protection and Digital Information Bill.} along with other legal provisions for digital identity solutions. The draft is still in discussion so far, and it is expected to consolidate the existing unrelated political initiatives and systems like "One Login"\footnote{Currently in Beta phase.}. This consolidation is expected to come though the Information Gateway\footnote{Section 54, Data Protection and Digital Information Bill.} that will allow trusted organizations to verify user's identity management data against the data held by public authorities. Coupling this with the ODIA and the certification processes for different organizations and digital ID solutions, the UK is getting closer to a national digital identity scheme.

\subsubsection{Identity Documents Act (Estonia)}
\label{sec:4:1:4}
The Identity Documents Act of 1999 made it mandatory for all Estonian residents to have an identity document. Electronic identity documents are just as well accepted as physical documents under this Act. We also note that the Estonian physical ID cards are equipped with a smart chip that contains private keys and can be used for online identification and authentication and even issuance of digital signatures, which means that national identity in Estonia is already a mix of the physical and digital components that make up the citizen’s complete identity. The Identity Document Act covers the provisions for the creation of identity documents, both digital and physical, and the personal data to be collected to issue them.

The Ministry of the Interior and the Police and Border Guard Board is responsible for the interpretation and enforcement of the Act. Responsible authorities are left to decide the list of certificates, identification and authentication procedures, the nature of the identity document database, and its governance. Giving executives more freedom to interpret the Act has made the implementation of the digital identity infrastructure and services more flexible. However, the legitimate purpose of the Act is to identify and authenticate Estonian residents, which means that the executive authorities can use the digital identity infrastructure as they see fit, since the use cases of the Act and the derived identity documents are not specified. 

The Identity Documents Act does not define the actors, but rather specifies the issuers of identity documents, such as the Police and Border Guard Board (PBGB), as the issuer of ID cards, which is also responsible for the legal identification of applicants and the management of the Identity Documents database. It is also important to mention that another Act, the Electronic Identification and Trust Services for Electronic Transactions Act, is more detailed and specific in defining ecosystem actors and trust lists, as well as policies for electronic identification and other electronic trust services such as digital signatures and seals.

Estonia is one of the most advanced countries in terms of experience with digital identity management, and beyond that, the digitization practices of public services. However, several major incidents have been reported in relation to Estonian digital identity solutions. Most significant is the 2017 Infineon incident, for which security flaws were detected in the chips of  approximately 750,000 ID cards issued between 2014 and 2017 \cite{Copper-Smith31} although the chips were subjected to a certification process \cite{FOIS32}.
This unfortunate experience raises concerns about the security of the digital infrastructure and actors in the digital identity ecosystem in Estonia, especially because the majority of actors are private sector companies.

Finally, the Identity Documents Act does not limit the purpose and scope of digital identity solutions and associated private data, as explained above. Complementary legislation, such as the GDPR, the Estonian Personal Data Protection Act (PDPA Estonia), The Electronic Identification and Trust Services for Electronic Transactions Act of 2016 and eIDAS regulation, provide more specific legislative guidance regarding the establishment of digital identity documents, the privacy and security of associated personal data, and the governance of the infrastructure and its use cases.

\subsubsection{Aadhaar Act (India)}
\label{sec:4:1:5}
The Targeted Delivery of Financial and Other Subsidies Benefits and Services (Aadhaar) Act was passed in 2016. It defines the legal framework for the Aadhaar identity program as an Act to provide effective, transparent, and targeted financial assistance and other services to individuals residing in India through the assignment of unique identity numbers. This identity program began in 2009 as a project to identify Indian residents.

Its adoption in 2016 was questionable. Indeed, the Aadhaar Act was adopted as a “money bill” as part of the budget presentation, with much criticism as it bypassed the upper house of parliament, since money bills must be passed only by the lower house. The identity project was also disputed in the Supreme Court, as the public authorities pushed for the use of the identifier in multiple use cases, in clear contradiction with the non-mandatory aspect of the Act \cite{India-SC33}. 

Beyond the questionable process of passing the Act, the Aadhaar Act and Aadhaar itself are the subject of much criticism. Even in Indian popular culture, Aadhaar identifiers have been greeted with great caution and scepticism due to the fear of increased government control. This is reflected in the famous example of the movie “aadhaar” released in 2019.
Moreover, the Aadhaar identity program, initially presented as optional, is finally made mandatory in practice. Indeed, Aadhaar IDs are required to access multiple subsidies and financial services, which are essential for the poor population. Enrolling a child into school also requires the provision of a child's Aadhaar ID.

Although the technical solution predates the Act itself, the Aadhaar Act legalizes the personal data collected during the registration and use of the Aadhaar platform. The Act defines the demographic and biometric data required to enroll in the national identity system, including name, date of birth, home address, gender and fingerprint, iris scan, and a photograph (as face recognition is supported by the solution). This personal data are stored in a central database called the Central Identities Data Repository (CIDR) and maintained by the UIDAI.

The Act also defines how registration agencies carry out the enrollment process, which means that it covers what data are to be collected, how, and by whom. During the data collection process, subjects must be informed of how the collected data will be used, with whom it will be shared, and the right they have to access their data and how to access them. India has not yet implemented a data protection legislation, as the proposed 2019 bill has been withdrawn \cite{TNYT-DP-India34}. Although there is another act, the Information Technology Act \cite{Aadhaar-text35}, that addresses some aspects and rules for data processing, it is positive to see that the Aadhaar Act sets rules for consent and user awareness for the national identity system.

The authentication process is also defined in the Act, which means that the legal framework effectively technically limits how the architecture of the solution is designed and how the authentication flow, as well as the registration flow, is realized. This is understandable, since the legislation has given a legal context to a solution that already exists, but it severely limits the solution on the technical side; since then, the legislation must be amended when new changes are introduced or different technological choices are made.
In the authentication process, the legislation also specifies that entities requesting the authentication of a user from the UIDAI must obtain user consent and inform the user of the purpose of the authentication, and the services and information the user will obtain after the authentication.

An authentication record, including the authentication time, requesting entity, and authentication response, is maintained by the UIDAI. This ensures a degree of accountability and traceability for the authenticating entities; however, it means that user activity can be easily traced, especially because the Aadhaar identifier is unique and persistent.
The demographic and biometric data required for the generation of Aadhaar IDs and deduplication are protected by law. Basic biometric data cannot be shared with any other party for any reason, is classified as “sensitive personal data” and falls under the provisions of the Information Technology Act, 2000 (21 of 2000) \cite{Aadhaar-text35}.

The Act gives UIDAI responsibility for the Aadhaar identity system and assigns operational and security duties to this authority.

\subsubsection{National Digital Identity Initiative Singapore}
\label{sec:4:1:6}
The National Digital Identity (NDI) is considered a digital enabler within the Digital Government Blueprint\footnote{Addendum to Digital Government Blueprint - 47.c (updated in 2020)} elaborated by the Government Technology Agency GovTech. The solution Singpass discussed above in the national digital identity solutions section is underpinned by the NDI. GovTech also appointed the National Certification Authority (mentioned when studying Singpass) and Assurity\footnote{\url{https://www.assurity.sg/}}. The first is responsible for the PKI and certificates behind Singpass, and the second provides different Singpass products like "Login" for onboarding Singpass users with high LoA. So overall, Singpass itself is not the result of a legislation but rather of a Digital Governance plan piloted by the Government Technology Agency of Singapore to provide reliable digital identity for both public and private sector. In terms of governance, the solution is maintained by GovTech and the entities it appoints like Assurity and the National Certification Authority. As for technical requirements and principles about personal data, it is GovTech that specifies those. For example, for private sector integration of Singpass products, technical requirements include the use of X.509 Public Key Certificates with RSA key sizes of 2048 bits and larger, provided by a compatible Certificate Authority\footnote{Can be found here \url{https://api.singpass.gov.sg/library/verify/developers/implementation-technical-requirements}}. As for personal data collection and processing requirements, the GovTech places personal data related to Singpass under the protection of the Personal Data Protection Act.

\subsubsection{Comparing Legal Frameworks}

As seen in the previous subsections, different legal frameworks have different legal purposes, requirements, and backgrounds for the solution, as depicted in Table \ref{tab:3}.
For example, some frameworks define limited use cases for their identity solutions (e.g., India, France), and others leave it open to interpretation by public authorities (e.g., Estonia). Some legal frameworks that contain certain technical specifications that influence how a solution is implemented (e.g., France and India) and others leave it up to the competent public authorities and entities to choose the technologies needed to implement the solution, as long as the legal requirements are met (e.g., Estonia and the eIDAS regulation).

\begin{table*}
\centering
\setlength{\tabcolsep}{8pt}
    \caption{Comparison of legal frameworks for national digital identity solutions.}
    \label{tab:3}
\begin{threeparttable}
\begin{tabular}{p{40pt}p{40pt}p{40pt} p{40pt}p{30pt}p{70pt}p{40pt}p{40pt}}
\hline
Legal Framework &
Adopted in &
Technical solution &
Supervisory body &
Year &
Legal purposes &
Technical requirements &
Other applicable legislation
\\
\hline
\\
eIDAS regulation &
EU Member States &
eIDAS notified solutions\tnote{1} &
European Commission &
2016&
Harmonize digital identity standards and requirements and ensure interoperability between member States &
None &
GDPR
\\
\\
The decree of November’s 8th&
France &
FranceConnect &
DINUM &
2018 &
Unify the digital identity infrastructure
to simplify online administrative procedures &
Federated Identity model &
GDPR,
Law on IT and Liberties
\\
\\
Data Protection and Digital Information Bill\tnote{2} &
UK &
Any solution trusted by the ODIA &
ODIA &
2023 &
Establish a legal framework for digital ID services and verification &
None &
Data Protection Act
\\
\\
Identity Documents Act &
Estonia &
eID Estonia &
PBGB &
1999 &
Identifying and authenticating residents &
None &
GDPR,
PDPA Estonia,
eIDAS regulation
\\
\\
Aadhaar Act &
India &
Aadhaar &
UIDAI &
2016 &
Delivery of subsidies and financial aid and services &
Centralized authentication process based on a central database CIDR &
Information Technology Act
\\
\\
National Digital Identity initiative &
Singapore &
Singpass &
GovTech, NCA &
2003 &
Provide reliable digital identity services for public and private sector &
X509 Certificates, RSA public keys \> 2048 bits, specific TLS ciphers &
Personal Data Protection Act
\\
\hline
    \end{tabular}

\begin{tablenotes}
\item[1] A list of currently eIDAS notified solutions can be found here \cite{eidas-notified36}
\item[2] The Bill is still a draft during the writing of this paper. 
\end{tablenotes}
\end{threeparttable}
\end{table*}

\subsection{Personal Data Protection Legislation}
\label{sec:4:2}
According to the United Nations, 137 countries already have data protection and privacy legislation in 2021. Including countries with a draft of such legislation, 80\% of States have such legislation \cite{UNCTAD37}. States without legislation and without any available data about such legislation sum up to 20\%, including India that we have studied in previous sections.
The definition of personal data and the policies in place to protect them, in addition to the various levels of regulation adoption, vary from State to State. 

We explore a few examples of legislation in this subsection: the GDPR and its consequent implementations applied in the EU; the Law on IT and Liberties (1978) in France, the Data Protection Act (2018) and the Data Protection and Digital Information (No. 2) Bill (2023) in the UK, Estonia's Personal Data Protection Act (2007/2018), The Personal Data Protection Act (2012) of Singapore and the California Consumer Protection Act (CCPA) (2018) as an example of personal data protection legislation in the United States.
We compared the legislation based on the definition of personal data, the rights granted to users, the identity of the protected subjects, and the location where the law applies.

\subsubsection{General Data Protection Regulation GDPR (EU)}
 The GDPR was published in the EU's official journal in 2016 and entered into effect in 2018 replacing the Data Protection Directive 95/46/EC of 1995. The regulation harmonizes data protection laws of EU Member States. It aims to protect individuals in the EU by defining personal data and rules on how data are collected and processed by private and public sector organizations, regardless of whether the organization itself is established in the EU.
 
The GDPR defines personal data as any information relating directly or indirectly to an identified or identifiable natural person. It defines the legal grounds for personal data processing and collection and cases where the data subject’s consent must be obtained. In matter of fact, GDPR requires personal data controllers and processors to have consent explicitly as a clear affirmation from an informed data subject that agrees to their data being processed\footnote{Article 4-11, GDPR.} in multiple cases.

GDPR gives the data subjects multiple rights, such as transparency on the part of the controller/processor of personal data about how data will be processed, by whom and with whom it may be shared, access to consult their collected data and correct it, the right to restrict the processing of the data, or to delete that data. Users also have the right to claim compensation when an organization causes any material or non-material losses because of non-respect of the regulation\footnote{Article 82, GDPR.}.

For their part, controllers and processors of personal data must comply with the GDPR principles such as fairness and transparency, purpose limitation, data minimization, accuracy, storage limitation and establishing measures to ensure the security, confidentiality, and integrity of personal data\footnote{Article 5, GDPR.}. Accountability is also a major principle, as controllers and processors of personal data must demonstrate compliance with the regulation, which legally require organizations to appoint a Data Protection Officer (DPO) if their primary activities include processing sensitive data on a large scale, although they may appoint a DPO voluntarily. The DPO’s role is to ensure that their organization(s) process personal data collected from employees, customers, and other possible data subjects in compliance with the GDPR and, potentially, other national laws dealing with data protection.

GDPR also requires each EU Member State to establish at least one Supervisory Authority to enforce GDPR\footnote{Chapter VI - Article 51, GDPR.}. These public independent authorities monitor the application of the regulation, and the application of other legislation derived from GDPR or based on it. In matter of fact, GDPR leaves to Member States certain flexibility in incorporating GDPR principles into their legal systems.
 
Given the size of the EU, with over 447 million people, and the extraterritorial nature of the GDPR, other States and organizations outside the EU have an interest to have GDPR-compliant legislation to stay in business with the vast European market\cite{BRADFORD}. Exporting data outside the EU is also possible but only to States with adequate levels of protection\footnote{Article 45, GDPR.}.
The UK for example kept the GDPR in place after Brexit. Other States such as Canada, Japan, and Argentina have an adequate level of protection. Therefore, transferring data to these States is expressly permitted \cite{GDPR-third-countries38}.

As part of the GDPR implementation, financial penalties of up to €20M are applied for non-compliance or incidents such as data breaches. These penalties account for up to 4\% of the organization's annual revenue if they exceed the €20M fine. The GDPR also explicitly requires organizations to report breaches to supervisory authorities within 72 hours of discovery, along with a justification for the delay. In terms of penalties, the GDPR is considered as very strict compared to other laws.

One of the strong aspects of GDPR is user awareness of the regulation, for both organizations that handle user personal data and users themselves. GDPR DPOs ensure the awareness of employees, and for the public, the survey requested by the European Commission about GDPR awareness \cite{Eurobarometer39} indicates that around 2/3 of EU residents have heard of the GDPR and 3/4 of them are aware of at least one right guaranteed by the regulation, with 1/3 of them aware of all rights.

\subsubsection{Act 78-17 on Data Processing, Data Files and Individual Liberties (France)}
The French Act 78-17 came as a response to the Safari affair mentioned above in the section summary of the \textbf{National digital identity solutions}. The attempt to use an unique identifier across public authorities and to inter-connect files held on individuals lead to the creation of a commission to provision measures and legislation to ensure the respect of private data and liberties. This later evolved to the CNIL that was established as an independent administrative authority by the Act 78-17 in the 6th of January 1978. This provided the ground for personal data protection in France. It was updated in 2004 to implement the 95/46/EC directive, and later updated in 2018 in the light of GDPR\footnote{Act 2018-493, 20 June 2018}. CNIL is designated as the personal data protection authority in France under GDPR.
Act 78-17 and GDPR are complementary, like most EU member State implementations of GDPR on a State level. In matter of fact, GDPR has empowered the CNIL in terms of sanctions and supervision since it gives the possibility to apply harder financial sanctions.

\subsubsection{The Data Protection Act and The Data Protection and Digital Information Bill (UK)}
The Data Protection Act (DPA) of 2018, also called the UK GDPR, is the UK's implementation of GDPR. DPA was passed to keep a GDPR-similar data protection legislation after the Brexit referendum of 2016. It retains GDPR notions and principles in the UK legal system with minor amendments for domestic context.
In matter of fact, up until the end of 2020, all personal data collected and processed were under GDPR, since GDPR was part of the retained EU laws that were incorporated to the UK legal system by the European Union (Withdrawal) Act 2018\footnote{An Act of the Parliament of the United Kingdom. Also known as the Great Repeal Act.}.
The DPA in its Part 5 - 114 and 115 - names The Information Commissioner as the UK's independent supervisory authority responsible of the Act application, as required by the article 51 of GDPR. UK GDPR also have an extra-territorial scope, the same as GDPR.

The UK Parliament is also currently discussing a new bill called The Data Protection and Digital Information (No. 2) Bill \cite{billNO240} that was first discussed in the committee stage in the house of Commons in March 2023\footnote{Note that we are interested in the V 2.0 of the bill since it is currently being discussed in the house of Commons. The first version was withdrawn on March 8th 2023.}. This new bill makes a few amendment to the already existing UK GDPR. In its Part 1 Data Protection, the bill changes some of the definitions and sections found in the Data Protection Act of 2018. For example, the bill substitutes some key GDPR terms found in the DPA with other terms, like "identifiable natural person" (Article 4- Definitions paragraph 1, GDPR.) that is replaced with "identifiable living individual" or "natural person" that is replaced with "individual". It also changes some paragraphs and notions like pseudonymisation and introduces new elements. For example, a DPO like the GDPR requires is no longer required in most cases, and a senior manager replaces this role in a context where the organization is involved in high-risk processing of personal data. Controllers or processors that are not established in the UK are no longer required to have representatives. It also allows data controllers to refuse a user request or charge a fee for performing it. Overall, the draft bill offers a softer version of the UK GDPR and of the GDPR, but still keeps a GDPR-compliant legislation.

Note that the Data Protection and Digital Information Bill does not replace the Data Protection Act of 2018. It rather amends it, and adds more legislation for digital identity in the UK. For instance, Part 3 of the Bill defines how the public authorities make provisions for the access to data and information about goods, services, digital content, supply and provision of such goods and services, customer feedback, transactions between customers and businesses etc.

\subsubsection{Personal Data Protection Act (Estonia)}

Estonia's Personal Data Protection Act\footnote{RT I 2007, 24, 127} was passed in 2007 and went into effect in 2008. It was since amended a few times by other legal instruments (like the Public Information Act\footnote{RT I 2000, 92, 597}).

Currently, the Personal Data Protection Act of 2018 which implements GDPR into Estonian law with the same scope of GDPR in terms of material and territorial scope. Although we notice that the Estonian Act extends the validity of the data subject's consent for 10 years after their death unless they decide otherwise and 20 years in the case of the death of a minor data subject (under 18 years old in Estonia)\footnote{Chapter 3 - Article 9, Personal Data Protection Act (2018).}.

In terms of definitions like the definition of personal data, health data and other GDPR definitions, the Act has no variation from the GDPR. The same goes for data subject rights.
However, although for legal bases of data processing both texts are very similar, the Personal Data Protection Act adds more ground for personal data processing like the transmission of personal data that is related to a violation of obligation\footnote{Chapter 3 - Article 10, Personal Data Protection Act (2018).} - like a contract breach by the data subject - to a third party for the purpose of the assessment of creditworthiness of the data subject and similar purposes, with exceptions for special categories of personal data like the ones under GDPR.

The Data Protection Inspectorate (DPI) is the data protection authority in Estonia. It is responsible for the application of both GDPR and the Personal Data Protection Act. The DPI is the extra-judicial party that handles sanctions defined by both texts\footnote{Chapter 6 - Liability Articles 62 to 73, Personal Data Protection Act (2018).}.

In matter of fact, GDPR penalties defined by Article 83 are not applicable in Estonia since the legal system of this State does not allow for administrative fines\footnote{See Recital 151, GDPR.}. It is the  Personal Data Protection Act that establishes the same sanctions in addition to other sanctions in the form of fine imposed by the DPI in the framework of a misdemeanour procedure.

\subsubsection{Personal Data Protection Act (SINGAPORE)}

The Personal Data Protection Act (PDPA) was passed in Singapore in 2012 and later revised in 2020. It regulates the collection, use, and disclosure of personal data for legitimate and reasonable purposes by organizations, and protects the personal data of individuals that are collected in Singapore. PDPA has an extraterritorial scope similar to GDPR, and it applies to overseas organizations that collect, use, or disclose data within Singapore.

To enforce and administrate the PDPA, a Personal Data Protection Commission (PDPC) has been established under the Info-communication Media Development Authority. However, the PDPA applies only to private sector companies, as the processing of personal data by public sector bodies is governed by another law called the Public Sector Governance Act 2018 (No. 5 of 2018). Moreover, all public-sector agencies manage data as a common resource shared by all entities within the public body.

The PDPA defines personal data as information about an individual who can be identified from this data or other data to which the organization has access to. Data related to business contact information, such as the name and position within a company, business phone number, and business email address, are excluded from this Act, and are not considered as personal data.

Like other personal data protection laws, the PDPA gives individuals the right to access and correct their personal data held by a company, and the right to see how that data has been processed. They also have the right to private action, which means that individuals who suffer loss or damage as a result of a breach of their personal data under this Act can file a complaint and sue the responsible organizations for compensation. As for consent, PDPA requires user consent for data collection, processing, and disclosure; however, there are multiple cases where consent is deemed given, such as consent by conduct (providing personal data for a specific purpose), by contractual necessity, and by notification \cite{PDPA-SG41}.

Companies must comply with the Act to avoid fines of up to 10\% of their annual turnover in Singapore or 1 million Singapore dollars (around €700,000), whichever is high. The 10\% fine is considerably higher than the 2\% or 4\% set by the GDPR. However, the percentages set by GDPR concern the company's global revenue and not just its local turnover like the PDPA specifies.

The rules on data collection and processing include limiting the purpose and scope of personal data processing and notifying the PDPC of any notifiable data breach. Data breaches are defined as breaches that result or may result in significant harm to an affected individual or if they are of significant magnitude. This means that unauthorized access to data, alteration of data or processing for other purposes, or disclosure of personal data are not notifiable breaches if they are performed on a small scale or for purposes that do not threaten individuals.

The transfer of personal data outside Singapore can only be done in accordance to requirements, like requiring the organizations to which the data are transferred to meet standards of protection comparable to those of the PDPA.

\subsubsection{California Consumer Privacy Act (USA)}
While the USA does not have a comprehensive federal personal data protection law, several States have bills in the drafting stage, and only a handful of States have already signed and implemented bills, such as California, Virginia, and Colorado. At the federal level, the USA has laws that address specific personal data protection, such as HIPAA, which addresses the privacy of health records, and FERPA, which addresses educational records, among others \cite{NYT-PLUSA42}.

The State of California passed the California Consumer Privacy Act (CCPA) in 2018, and it went into effect in 2020. The Act establishes rights for California residents related to their personal data. Personal data is defined as any information that identifies or can be used to identify a natural person or their household. This definition is different from that found in the GDPR, which focuses on protecting individual personal data, unlike the CCPA, which extends that protection to household data, including data from home devices. Moreover, the CCPA protects residents of California in the State and even outside of it when they are temporarily away, but does not apply on visitors of the State. CCPA only applies to private for-profit companies that meet certain criteria, such as their annual revenue (over \$25M or at least 50\% of it comes from the sale of personal data) and their number of customers (over 50,000). The CCPA does not apply to public authorities or non-profit organizations.

The rights established by the CCPA include those similar to those defined in the GDPR. For example, the right to transparency regarding access to data collected and held about them, the right to prevent companies from selling their data to third parties, and the right to delete data. However, the CCPA does not require consent from the user before collecting and processing their data. CCPA does not always require organizations to obtain the user's consent with an opt-in form like GDPR does in most cases, so it introduces the right to opt-out\footnote{Section 5 - 1798.120. (a), CCPA.}, allowing users to withdraw their default consent.
CCPA requires consent in "opt-in" form only in one specific case: the sale personal data of a minor (under 16 of age)\footnote{Section 5 - 1798.120. (c), CCPA.}. This requires affirmative authorization of the legal guardien of the minor to sell such data.

Unlike the GDPR, the CCPA does not include fines or penalties for non-compliance. Instead, these penalties were applied in the case of a data breach. The maximum amounts of these fines are very low compared to the GDPR, for example, ranging from \$2,500 to \$7,500 for each breach, a sum that can be considerable if there are many breaches. This makes the CCPA very reactive, as it does not preemptively sanction non-compliant organizations. It is also worth noting that only data breaches with limited circumstances under the CCPA are eligible for legal action by a user, such as when a company fails to take reasonable security measures to protect user's data. Most other breaches are not actionable, and only the Attorney General can bring lawsuits to companies.

CCPA was amended by the California Privacy Rights Act (CPRA) of 2020 that was voted on November 2020 ballot. CPRA passed in 2020 and went into effect - in full - in January 2023. The new Act adds more privacy protection to CCPA and notably establishes the California Privacy Protection Agency\footnote{Section 24 - 1798.199,10. (a), CPRA.} to enforce CPRA and CCPA alongside the Attorney General that also retain the civil enforcement authority.

\subsubsection{Section Summary}
As seen in this subsection, data protection laws around the world have been spurred by the requirements and extraterritorial effect that the GDPR imposes on entities that wish to continue offering services in the European market. Since the introduction of the GDPR, many countries have sought to introduce their own laws or update their laws to be compatible with the GDPR.

It is positive to see that personal data protection Acts are being adopted by countries even though they do not have national digital identity solutions. From one end, such Acts will provide protection for users while using any online solutions. From another end, they will lay the appropriate legislative ground for future digital identity systems.

However, these laws have different definitions of personal data, scope, and requirements imposed on data controllers and processors, just as they are different in terms of what rights they grant to the entities they protect. As discussed before, they also have differences in the entities covered (natural persons, legal persons...) and the organizations that are concerned (private sector entities, public sector entities…).  Table \ref{tab:4} depicts this.

Personal data protection acts need some sort of harmonization, especially in a context where data can be collected in one State and processed in another or where data simply move between countries. At another level, the differences between these acts sometimes make it difficult for legal frameworks – and therefore, solutions – to be interoperable.

\begin{table*}
\centering
\setlength{\tabcolsep}{7pt}
    \caption{Comparison between data protection laws.}
    \label{tab:4}
\begin{threeparttable}
\begin{tabular}{p{60pt}p{60pt}p{70pt}p{60pt}p{60pt}p{60pt}}
\hline
Legislation &
Adopted in &
Year &
Supervisory Body &
Scope &
Fines and penalties\tnote{1}
\\ 
\hline
\\
GDPR &
EU &
2016 passed,
2018 effective &
EU: European Data Protection Board
Member State: Data Protection Authority\tnote{2} &
Public and Private Sector organizations
&
2\% or up to €10M for less severe offenses
4\% or up to €20M for serious offenses
\\
\\
Act 78-17 on Data Processing, Data Files and Individual Liberties &
France &
1978 passed,
2004, 2018 major updates
&
CNIL
&
Public and Private Sector organizations
&
Same fines as GDPR,
Temporary or permanent limitation of data processing,
certificate revocation, etc

\\
\\
The Data Protection Act and The Data Protection and Digital Information (No. 2) Bill &
UK &
DPA in 2018, while the bill is still in discussion
&
The Information Commissioner (ICO) Office
&
Public and Private Sector organizations, inside and oustide of the UK
&
£8.7 million or 2\% as standard maximum,
£17.5 million or 4\% as higher maximum
\\
\\
Personal Data Protection Act &
Estonia &
2007 passed,
2008 effective,
2018 implemented GDPR
&
DPI
&
Public and Private Sector organizations
&
Same as GDPR, but enforced by the DPI as extra-judicial misdemeanour procedure
\\
\\
Personal Data Protection Act &
Singapore &
2012 passed,
2020 revised &
PDPC &
Private Sector organizations, Personal data of individuals that are collected in Singapore, processed by a private entity and any entity acting on their behalf &
10\% of turnover in Singapore or up to 1 million Singapore Dollar
\\
\\
CCPA &
California, USA &
2018 passed,
2020 effective &
California Privacy Protection Agency &
Private Sector and for profit companies that satisfy certain conditions, California residents and their households devices &
Between \$2,500 and \$7,500 per breach
Between \$100 and \$750 as compensation per individual per incident
\\ \\
\hline
    \end{tabular}

\begin{tablenotes}
\item[1] Percentages are from the annual turnover of the organizations that are fined, unless stated otherwise.
\item[2] Each EU member State has its own Personal Data Protection Authority. The list of authorities per State can be found here \cite{edpb-list43}

\end{tablenotes}
\end{threeparttable}
\end{table*}

\section{Authors Position}
\label{sec:5}
In a world where State surveillance and cyber-crime are on the rise \cite{survstat44} and where the need for trust by all actors, from States to citizens to service providers, is keenly felt, we believe that SSI solutions offer the best technical and governance conditions for building a national digital identity management system. SSI solutions operate in zero-trust environments where the trust is rooted on decentralized ledgers, users have full control and ownership of their identity management data and able to present service providers with reliable credentials certified by trusted issuers.

We also believe that SSI solutions should not eliminate the role of public authorities and trust service providers, but should integrate them into the SSI ecosystem in an effective way that balances the user’s experience, security, and liability, and the efficiency of a national identity system. Moreover, we assert that a national digital identity solution should be able to provide a trusted way for businesses and private entities to onboard\footnote{Onboarding is the whole process done by a business to enroll a customer, have them activate the service or the product and use it. This includes the identification and authentication of users in a continuous way.} their customers and for customers to enroll and subscribe to different services in a simple, secure, and privacy-preserving manner.

\subsection{Blockchain-based SSI as a National Digital Identity Solution}
\label{sec:5:1}

An SSI solution revolves around two major components: the wallet application that enables users and stakeholders to interact and exchange identity management data, and the blockchain infrastructure that acts as a decentralized root of trust. Implementing a national digital identity solution that follows an SSI architecture and based on a blockchain infrastructure requires careful planning and configuration for both major components.

\subsubsection{Wallet Applications}
Wallet applications are a digital version of the systems we already have and use: physical wallets containing government-issued identity documents and other types of documents that users collect. The close analogy between digital and physical wallets makes it intuitive for users.

Although other identity management models can rely on a wallet solution, the dependence on IdPs to generate identifiers for users limits the autonomy of users and undermines the concept of user control. Therefore, we believe wallets within the SSI model have more advantages than other types of wallets.

As with traditional physical wallets, users cannot be expected to carry multiple wallets for different purposes. Therefore, we strongly emphasize that regardless of the wallet provider, all wallets should have common core functionalities that allow users to hold and present different credentials from different issuers to different service providers. Wallets should be able to work with different existing SSI infrastructures while providing a seamless experience for the user. A user should be able to decide managing his IDs through several independent wallets, e.g. one for each specific purpose, or one wallet for all purposes. If a user has a wallet, they should be able to use it to interact with different issuers and systems, even if they belong to different SSI solutions. Similarly, they should be able to present the credentials held within the same wallet in a reliable, secure, and verifiable manner to different service providers, even if they do not belong to the same SSI ecosystem.

As for portability, users should be able to transfer their credentials from one wallet solution to another, and credentials should remain verifiable across a multitude of solutions and infrastructures. In addition to technical interoperability, there must be legal interoperability, meaning that issuers and wallet providers must adhere to a high level of standardization and certification so that credentials can be stored, recognized, and verified across different platforms and different wallet solutions.

Moreover, wallets must be provided by certified trusted wallet providers, which are held to high standards with respect to how the wallet is developed and maintained, its functionality, the device on which it is installed, and further considerations. The eIDAS Expert Group’s draft EUDI wallet Architecture and Reference Framework \cite{ARF-eIDAS23}, for example, provides multiple technical requirements that define how a wallet should operate and in what ecosystem. It defines roles in the ecosystem through a list of functional and non-functional requirements for wallets that are intended to be used in the EU in different SSI solutions. These requirements are being considered while defining the eIDAS 2.0 regulation, that will probably reflect them in a less technologically-neutral legislation.

Although we generally have a preference for more technologically-neutral legislation for fear of limitations and partial exploitation of technological capacities and innovation, in the case of SSI wallets that are used as national identity schemes and are intertwined with official documents, we believe it is reasonable that the legislation defines the limits of the technical solutions. This will gradually help citizens to get secure and portable SSI solutions. Accurately defining the entities involved in an SSI ecosystem and their role is essential for regulating and supervising SSI solutions.

We believe that an SSI solution should be introduced and adopted gradually by limiting the use cases. For example, we find it reasonable that ESSIF, the European SSI Framework initiative \cite{ESSIF-lab46}, starts with a single, well-defined use case, namely the cross-border recognition of educational credentials between EU Member States. This use case will enable a more dynamic movement within the EU, since it will drastically reduce certificate fraud and background check costs of recruiting and movement for potential employers and EU students. Limiting the initial scope of an SSI solution limits the parties involved, such as issuers and verifiers, making it easier to regulate and enforce the legal framework within which the solution operates.  This initial limitation will help test SSI solutions and their wallets with regard to usability and resistance to security attacks; however, it will also limit how wallets and SSI solutions used in different use cases can interoperate.

For SSI, wallet providers and the full SSI ecosystem must have significant security and privacy guarantees. For instance, when a wallet is implemented through a mobile application, the analogy with traditional physical wallets fails miserably. A physical wallet is in the total control of its owner, whereas a mobile application is hosted on a device. This means that the mobile manufacturer, operating system creator, app store, mobile operator, and wallet provider (the wallet application developer) must be trusted to ensure that users control their wallets and credentials. This requires that most involved parties are certified and held to high requirements and specifications on how a wallet should be developed, how a hosting device should be manufactured, and the technical requirements that a device should meet (NFC capabilities, Secure Environment SE, etc.).

From a personal data protection perspective, wallet providers should not be able to collect or access personal data stored on wallets. That is, in the case of multiple identities managed by the same wallet, the wallet implementation must ensure users that their identities cannot be correlated by the entities they interact with, for example, service providers and issuers.

Moreover, it is essential that users have the choice - whenever it is possible - to use pseudonyms and present themselves under the identity they wish \cite{Claire-LB47}. This means that wallets should implement privacy-preserving methods like Zero-Knowledge Proofs, selective disclosure and the capacity to generate peer-DIDs and different DID methods. Of course, their pseudonyms must not be linked by entities to their real ID, and the wallet can still provide the ability to prove non-identifiable personal data, such as age, license possession, and State of residence, satisfying certain age conditions using cryptographic techniques such as zero knowledge proofs.

\subsubsection{Blockchain Infrastructure}

As the previous sub-section was user-focused, to ensure a seamless user experience empowered by user control and security with privacy guarantees, we must not forget that a digital identity infrastructure is a matter of State sovereignty when it is used to access public services and essential services like energy and help identify and authenticate citizens with official government-issued identity credentials. These sovereignty requirements impose certain conditions and measures to be implemented by the public authorities.

On the other hand, national digital identity solutions that do not follow the SSI model heavily rely on the IdPs. Whether the IdP is a public authority or a delegated one, the trust model is centralized. Tipping the scale to the favor of users requires a paradigm-shift, that can be achieved through the use of a decentralized infrastructure. We argue that this infrastructure should be a consortium blockchain.

Maintaining DID identifiers truly owned by their users implies the need for a public Blockchain or registry. This ensures that no single entity can claim registry ownership, and by transitioning the ownership of the identifiers themselves, as in other identity models. However, to have reliable infrastructure for national identity credentials, we cannot expect the use of a public Blockchain because of the need for more trusted nodes and availability. We believe that a consortium Blockchain, operated by different public authorities and entities acting on their behalf, with the inclusion of non-governmental organizations to ensure openness and transparency, even if they only have an observatory role, is a more sensible choice.

We argue that a consortium blockchain offers more sovereignty to both public authorities and users since the infrastructure will not be controlled or managed by a single entity. A decentralized blockchain ensures that the public authorities are part of the infrastructure and that the no single entity of the other involved parties own the infrastructure and single-handedly control it. This removes the opacity between the public authorities and the digital identity used to access the services they provide, and ensures that the system’s infrastructure remains available even if some partners are not available.

\subsection{State sovereignty over the SSI Ecosystem} 
\label{sec:5:2}
Public authorities play a crucial role in a national SSI ecosystem. 
They must regulate the ecosystem with respect to national identity schemes and the legal framework. Sovereignty implies establishing a comprehensive certification methodology carried out by a designated entity acting on behalf of a public authority (for example, Conformity Assessment Bodies in eIDAS 2.0).
The public authorities mediate between different issuers (public authorities, universities, and trust service providers), holders (mainly citizens and residents), and verifiers (service providers or entities acting on their behalf). Moreover, controls should be put in place to verify that service providers are complying with regulation. For example, service providers should not ask users to authenticate with a higher level of assurance than what is adequate, relevant, and necessary for the service offered (in accordance with the data minimization principle found in multiple legislation like GDRP, CCPA and others).

Public authorities also play an important role as issuers of official digital identity credentials, such as passports, driving licenses, identity cards (where they are used), and other forms of government-issued credentials. Rooting these credentials (in terms of verifiability like published hashes, publishing the signature keys, the revocation registries etc) on a consortium blockchain ensures that they remain available and that the authorities still control their revocation, without relying on any third-party infrastructure or privately owned ones.
Relying on external or private partners can lead to the breakdown of a national identity solution. For example, the UK Verify national identity scheme failed due to over-reliance on private entities, among other factors. 

This is valid for the choice of wallet providers, the underlying technological solutions (e.g., mobile manufacturer), the choice of electronic ID issuers, and the trust services providers. Preference must be given to national companies that are already associated with a State's activities to offer better guarantees of long-term sovereignty. 

Because the interaction occurs online, public authorities should publish trusted lists and other registries used to recognize and certify trust entities, such as issuers, verifiers, and trust service providers. Indeed, the validity and trust of an SSI system directly depends on the validity and trust of its issuers and the credentials they offer. SSI systems begin with empty wallets, and trust is built through the exchange of credentials and a mutual authentication procedure. Trust is also built over qualified trust services, which brings complementary electronic evidence to stakeholders that can be used in court. Public authorities and Trust Service Providers remain an important trust anchor in SSI systems, and they are considered assets because they represent sources of trust.

In the EU dynamics, for example, adopting the digital identity wallets is a paradigm shift that will require EU Member States to issue wallet applications under their eIDAS notified digital identity solutions \cite{ECPR48}. This shift came with new provisions - introduced by the proposal of regulation eIDAS 2.0 - about wallet certification and secure storage of cryptographic materials. Such provision are put in place to ensure trust in wallet solutions.

Establishing a fully functional SSI solution, with a strong infrastructure and trusted wallets, will enable citizens and residents to identify and authenticate themselves securely on different Levels of Assurance (LoA) for different services. This will make the onboarding process faster for both private and public services, and users will no longer be required to go through a third party to digitally sign documents and contracts since they can directly sign with keys associated to their DID identifiers. There will no longer be the tedious process of taking scans and photos of physical identity documents along with selfies and check-in phone calls to verify an identity. An SSI wallet will enable a user to select whatever attributes are required from their existing credentials and securely present them to a verifier to access online and offline services.

This opens up a lot of opportunities for public-private partnerships, where public authorities can offer secure onboarding and authenticated identities, enabling faster and secure transactions between citizens and businesses.

\subsection{SSI as an enabler of public-private partnership}
\label{sec:5:3}
SSI systems are easier to open to the private sector than other systems because of their decentralized nature and the high level of user autonomy they provide. We argue that the same government-issued wallet can be used and accepted by private companies if they set up their own agents to verify and exchange credentials with users and that an existing SSI infrastructure can still work with privately issued wallets.

Generally, public authorities lack the industrial and technological experience to develop, deploy, and pilot a national technical platform that offers relevant and scalable services to citizens and other stakeholders. Therefore, in most cases, a partnership based on delegation to private sector companies is inevitable. Private entities already play an important role in most national identity infrastructures by providing technology (for example, Guardtime in Estonia and Infineon, which integrates smart chips into a wide variety of e-ID cards), or by acting as identity providers for State-sponsored digital identity solutions (La Poste and Orange as IdPs for FranceConnect, Verizon, and Experian as IdPs for UK Verify).

States must encourage industries to create value and services around national digital identity solutions to ensure that the ecosystem is sustainable. Most existing private partners can carry on playing similar roles in a State-sponsored SSI solution: partners providing identity software solutions already have a head start and experience to work on identity wallets as wallet providers; entities acting as identity providers can benefit from their established systems and certifications and act as issuers, verifiers, and trust service providers; entities providing hardware and infrastructure solutions would maintain a crucial role in providing technology and support for the wide decentralized network that an SSI needs (Blockchain nodes, back-end servers..).

However, this partnership should not be at the cost of data protection abuse through abusive data collection. States are responsible for strict supervision of activities related to the implementation and operation of national digital identity systems.

Private companies can benefit from an SSI ecosystem in several ways. According to several reports \cite{ENISA-report49}, enrolling users and remote identity proofing have been a real challenge for many organizations. This was highlighted during the Covid-19 pandemic when most services shifted online. From a business perspective, online client acquisition involves various levels of identification and authentication. Whether there are high LoA requirements for banks, mobile operators, public services, and insurance companies or lower LoA levels for other businesses, enrollment, and verification of user identity and claims are persistent needs. The ability to identify, authenticate, and verify presented claims in a secure and reliable manner is a growing need for both public and private sectors. We believe that SSI can contribute to high standards and an efficient means of meeting this need, and can also reduce the costs.

In the SSI ecosystem, not only different issuers provide different credentials with different LoAs, but also users can select and combine attributes from different verifiable credentials coming from different issuers. This creates a wide variety of possible verifiable credentials that can be presented to service providers and verifiers that satisfy the required information. This comes with privacy insurance, such as providing information without disclosing it (zero-knowledge proofs), for example, that a user is of legal age without sharing their real age or birthdate, that a user is a resident in a certain State without sharing their home address, etc. In addition, security methods are implemented with wallet capabilities, such as electronic signatures and electronic seals. This proof can be obtained both online and offline in a trusted secure verifiable way.

This changes remote onboarding and identity proofing: users are no longer required to take photos of themselves and an official identity document to prove their identity, as they can directly present a verifiable digital version of those identity documents. This relieves the service provider of the process of verifying official documents and collecting personal data, and puts responsibility on the public authorities to give correct identification and asserted claims to users, while giving them the freedom to create their own DID identifiers and manage their authentication material (keys) through their wallets.

The costs of Know Your Customer\footnote{The mandatory check done by financial institutions to identify and verify the identity of their customers when onboarding them and later on regularly over time.} (KYC) are variable and depend on multiple factors from one State to another, and from one business to another. The KYC is mandatory for some businesses, such as banks and insurance, and very useful for other businesses, even if it is not mandatory. However, KYC procedures, which include identity verification and authentication and dealing with customer data, are expensive. Reports indicate that for banks, KYC costs a single bank approximately \$60M a year \cite{KYC-cost50} with approximately \$10.4B paid in fines for non-compliance or failure to correctly identify and authenticate customers worldwide \cite{banking-KYC51}.
SSI systems can drastically reduce these costs by providing users with secure means of identity-proofing and providing businesses with verifiable credentials that attest to the user’s identity.

\section{Conclusion}
\label{sec:6}
Today, we are at the crossroads when it comes to digital identities, whether it is establishing a system of mass-surveillance by States and/or companies or a digitally secure system that respects the freedoms of citizens to present themselves in any identity they wish. Of course, we favor the latter.

The ultimate goal is to restore trust in digital technology, allowing companies to do business; citizens to carry out their personal, administrative, and professional activities; and Sates to provide a protective regulatory framework for all stakeholders. The objective is to prevent fraud and data plundering by any actor, including the wallet providers. Beyond that, it is about preventing some actors from taking control of the digital world through digital identities. Let us not forget that the management of citizens' identities is a matter of the State, which must not be entirely delegated to the private sector; otherwise, the States risk being deprived of one of their essential prerogatives.

For this, it is necessary to set up a fair and healthy digital identity system where the States and the citizens keep control of the digital identities; that is to say, for the first one, the creation and revocation of the digital identities in the system with all the necessary regulation, standardization, and product and entity certification, and for the latter, having the choice to decide on the private actors with whom to interact and to entrust the storage and management of their identity data. We believe that SSI is the solution that best meets the objectives of a fair and healthy system. 

The authors hope that this position paper will help policymakers make informed decisions about national digital identity systems, and will help system designers and developers make better technical choices, as this system is a major societal issue that will determine the entire relationship between citizens and the digital world, a relationship of obligation and reluctance, or a relationship of cross confidence. This will determine whether citizens systematically adopt digital technology in their daily interactions or turn away from it whenever they can.

\section{Acknowledgment}
We thank Jonathan Keller, postdoctoral researcher in Law at Télécom Paris and member of both Chair Connected Car and Cybersecurity (C3S) and Chair Values and Policies of Personal Information for his valuable feedback during the writing of the paper.
This paper is partly supported by the chair Values and Policies of Personal Information, Institut Mines-Telecom, France, and European Union’s PRIMA Research and Innovation Action 2022 under grant agreement No XXX, project MoreMedDiet.

\end{document}